\documentclass[aps, prl, twocolumn,showkeys,showpacs,amsmath,amssymb,superscriptaddress]{revtex4-1}

\usepackage{lineno,hyperref}
\modulolinenumbers[5]

\usepackage{todonotes}
\usepackage{makeidx}
\usepackage{graphics}
\usepackage{color}
\usepackage{listings}
\usepackage{bm}
\usepackage{url}
\usepackage{booktabs}
\usepackage{amsmath}
\usepackage{paralist}
\usepackage{amssymb}
\usepackage{amsthm}
\usepackage{subcaption}

\graphicspath{ {img/} {plot/} }
\newcommand*\diff{\mathop{}\!\mathrm{d}}

\begin{document}


\title{Kinetic approach to relativistic dissipation}

\author{A. Gabbana}
\affiliation{Universit\`a di Ferrara and INFN-Ferrara, 
             Via Saragat 1, I-44122 Ferrara, Italy.}

\author{M. Mendoza}
\affiliation{ETH Z\"urich, Computational Physics for Engineering Materials, 
             Institute for Building Materials, Schafmattstra{\ss}e 6, HIF, 
             CH-8093 Z\"urich, Switzerland.}

\author{S. Succi}
\affiliation{Istituto per le Applicazioni del Calcolo C.N.R., 
             Via dei Taurini, 19 00185 Rome, Italy, and\\
             Institute for Applied Computational Science, John Paulson School of Engineering and Applied Sciences, Harvard University, Cambridge, USA}

\author{R. Tripiccione}
\affiliation{Universit\`a di Ferrara and INFN-Ferrara, 
             Via Saragat 1, I-44122 Ferrara, Italy.}

\begin{abstract}
  Despite a long record of intense efforts, the basic mechanisms by which dissipation emerges from the microscopic 
  dynamics of a relativistic fluid still elude a complete understanding. 
  In particular, several details must still be finalized in the pathway from kinetic theory to hydrodynamics mainly in the derivation of the values of the transport coefficients.
  In this Letter, we approach the problem by matching data from lattice kinetic simulations with analytical predictions.
  Our numerical results provide neat evidence in favour of the Chapman-Enskog procedure, as suggested by
  recently theoretical analyses, along with qualitative hints at the 
  basic reasons why the Chapman-Enskog expansion might be better suited than Grad's method to capture the 
  emergence of dissipative effects in relativistic fluids.  
\end{abstract}

\maketitle
 

The basic mechanisms by which dissipative effects emerge from the microscopic 
dynamics of relativistic fluids remains are still not fully understood in relativistic hydrodynamics.
It has been long-recognized that the parabolic nature of the  Laplace operator is
inconsistent with relativistic invariance, as it implies superluminal propagation, hence 
non-causal and unstable behavior \cite{hiscock-1983, hiscock-1985, hiscock-1987}. 
This can be corrected by resorting to fully-hyperbolic formulations of relativistic hydrodynamics, whereby
space and time come on the same first-order footing, but the exact form of the 
resulting equations is not uniquely fixed by macroscopic symmetry arguments and thus remains open to debate.

A more fundamental approach is to derive relativistic hydrodynamics from the 
underlying kinetic theory \cite{degroot-1980}, exploiting the advantages of the bottom-up approach:
irreversibility is encoded within a local H-theorem \cite{cercignani-2002}, while dissipation results
as an emergent manifestation of weak departure from local equilibrium (low Knudsen-number assumption)
and the consequent enslaving of the fast modes to the slow hydrodynamic ones, associated with 
microscopic conservation laws.  At no point does this scenario involve second order derivatives 
in space, thus preserving relativistic invariance by construction.
\begin{figure}[htb]
  \centering
  \includegraphics[width=.45\textwidth]{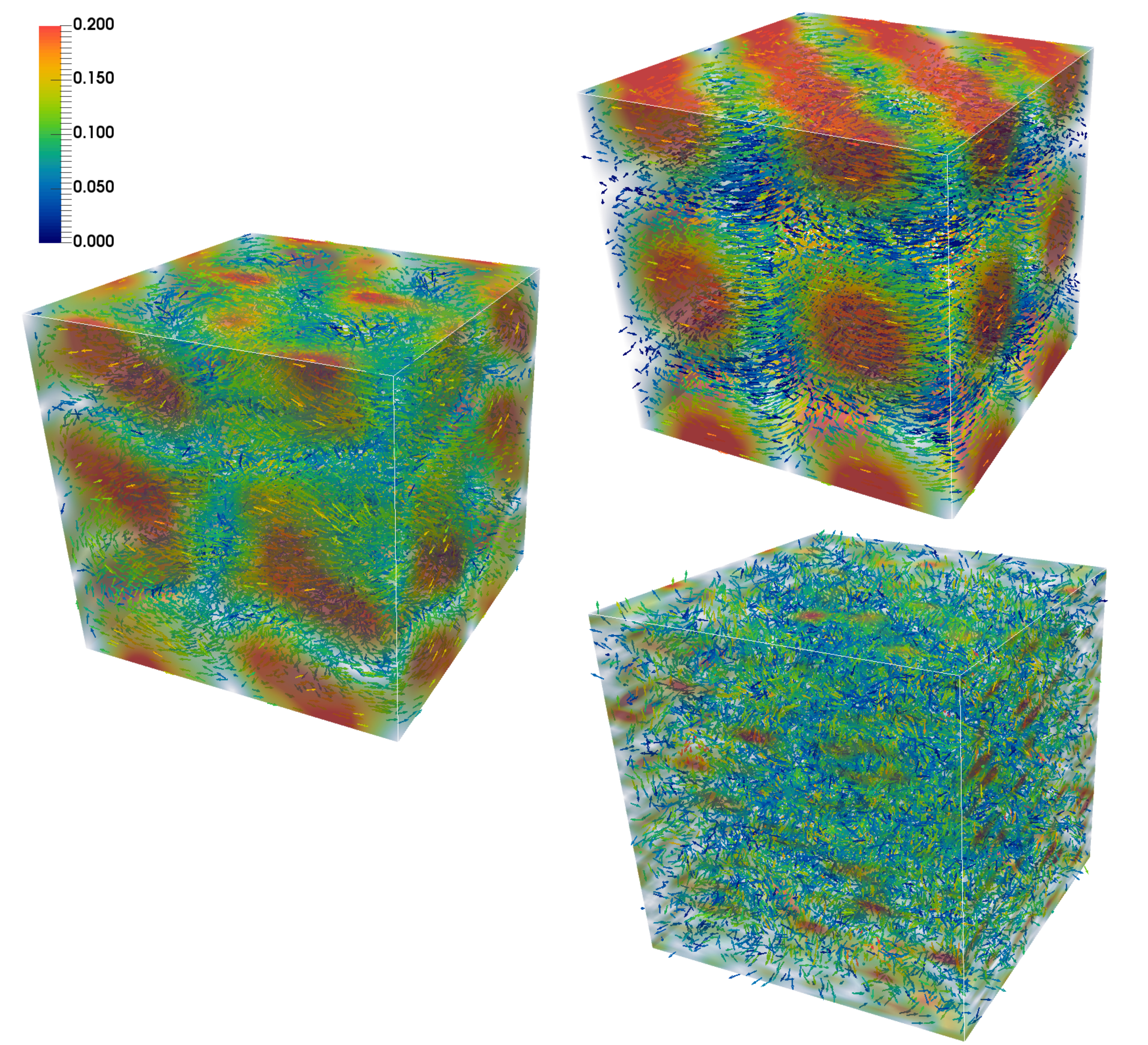}
  \caption{Three-dimensional Taylor-Green vortex configuration of a viscous relativistic fluid,
  with $\zeta = 0$ and $\tau = 0.51$ (symbols defined in the text). 
  Top: initial configuration; Middle: later stage in which the vortex configuration becomes unstable;
  Bottom: final disordered state. 
  Colours code vorticity and arrows represent the velocity field.}\label{fig:gtaylor3D}
\end{figure}  

In non-relativistic regimes, Grad's moments method \cite{grad-1949}
and the Chapman-Enskog (CE) \cite{chapman-1970} approach manage to connect 
kinetic theory and hydrodynamics in a consistent way, i.e. they provide the same
transport coefficients. 
However, the relativistic regime presents a more controversial picture.
The Israel and Stewards (IS) formulation \cite{israel-1976b, israel-1979}, extending Grad's
method, derives causal and stable equations of motion, at least
for hydrodynamics regimes \cite{romatschke-2010}. 
While many earlier works have relied on IS, recent developments have highlighted
theoretical shortcomings \cite{denicol-2012} and poor agreement with
numerical solutions of the Boltzmann equation \cite{huovinen-2009, bouras-2010}.

Recently, several authors have developed new attempts to derive consistent relativistic dissipative hydrodynamics equations.
Attempting to circumvent the drawbacks of the IS formulation, Denicol et al. \cite{denicol-2010, denicol-2012, molnar-2014} 
have proposed an extension of the moments methods in which the resulting equations of motion are derived directly from the
Boltzmann equation and truncated by a systematic power-counting scheme in Knudsen number. 

This, in turn, offers the possibility to include a larger number of moments (with respect to the 14 used in the IS formulation), 
improving the expressions for the transport coefficients.
Starting from similar considerations, Jaiswal et al. \cite{jaiswal-2013} have included entropic arguments within Grad's method and 
derived relativistic dissipative hydrodynamics equations which take the same form as IS, although with 
different expressions for the transport coefficients.
When compared to IS, these developments lead to solutions closer to the Boltzmann equation and,
at least in the ultra-relativistic limit (defined by $\zeta \rightarrow 0$, where $\zeta = mc^2/K_B~T$ is the ratio of particle
rest energy and temperature), they yield transport coefficients in good agreement with those 
calculated via the CE expansion.
Interestingly, the CE method itself remains somewhat less explored \cite{jaiswal-2013a, jaiswal-2013b}, with 
relativistic extensions mostly restricted to the relaxation time approximation. 
More recently, a novel approach, introduced in a series of works by Tsumura et al. \cite{tsumura-2012,tsumura-2015,kikuchi-2015,kikuchi-2016}, applies 
renormalization group techniques to the Boltzmann equation.
Once again, expressions for bulk (shear) viscosity and heat conductivity coincide with
those provided by the CE method.
Summing up, the present and somewhat not fully conclusive state of affairs, is that different theoretical 
approaches, based on different, if not conflicting assumptions, seem to converge towards the results provided by the CE approach.
Conceptual shortcomings of the moments method, recently highlighted also in the non-relativistic framework 
\cite{velasco-2002, struchtrup-2003, ottinger-2010, torrilhon-2016}, revolve around the use of second-order spatial derivatives 
in constitutive hydrodynamical equations \cite{tsumura-2012}. On the other hand, objections to the relativistic Chapman-Enskog expansion 
point to its link to relativistic Navier-Stokes equations, which suffer of basic problems, such as broken causality and resulting instabilities 
\cite{denicol-2010,denicol-2012}.
In a less than crystal-clear situation, one would like to validate theory
towards experimental data, but a controlled experimental setup is not a viable
option at this point in time. Given the circumstances, numerical simulation stands 
up as a very precious alternative to gain new insights into this problem.

Recent works \cite{florkowski-2013,
bhalerao-2014} have presented 1D simulations of the
(ultra)-relativistic Boltzmann equation in the relaxation time  approximation,
showing results asymptotically compatible with the CE approach. 
This letter follows a similar line and reports the results of lattice-kinetic simulations of a
relativistic flow in a controlled setup for which an approximate analytical
hydrodynamic solution can be derived.  
We match analytical and numerical results in order to
study the dependence of hydrodynamic transport coefficients on parameters
defined at the mesoscale. 
To this purpose, we study the time evolution of a Taylor-Green vortex
configuration in two and three spatial dimensions (see \autoref{fig:gtaylor3D})
and probe the functional dependence  of the transport coefficients on
$\zeta$, extending previous work confined to the $\zeta \to 0$ limit. 
Our main result is a neat indication that CE predictions accurately
match numerical data, and they do so over a remarkably wide $\zeta$ range, starting
from the ultra-relativistic regime and seamlessly going over to the well-known
non relativistic case.
Our simulations use a recently developed relativistic lattice Boltzmann
model (RLBM) \cite{gabbana-2017}, able to handle massive particles,
providing, to the best of our knowledge, the first analysis  of dissipative
effects for relativistic, but not-necessarily ultra-relativistic, flows.


In relativistic fluid dynamics, ideal non-degenerate fluids are described by the particle four-flow 
and energy momentum tensors, which at equilibrium read:
\begin{align}
  N^{\alpha}_E       = & ~ n U^{\alpha}                                                \quad , \\
  T^{\alpha \beta}_E = & ~ (P + \epsilon) U^{\alpha} U^{\beta} - P g^{\alpha \beta  }  \quad ,
\end{align}
where $U^{\alpha} = \gamma ~ ( 1, \bm{u} )$ is the fluid four velocity, ( $\bm{u}$ is the fluid velocity, $\gamma = 1/ \sqrt{1-u^2}$; we use natural 
units such that $c = 1,~ K_B = 1$), $P$ the hydrostatic pressure, and $\epsilon$~($n$) energy~(particle) density. 
We take into account dissipative effects with the Landau-Lifshitz decomposition \cite{cercignani-2002}:
\begin{align}
  N^{\alpha}       &= N^{\alpha}_E - \frac{n}{P + \epsilon} q^{\alpha} \quad \label{eq:1st-order-tensor} \quad , \\
  T^{\alpha \beta} &= T^{\alpha \beta}_E + 
                      P^{< \alpha \beta >} - \varpi \left( g^{\alpha \beta} - U^{\alpha} U^{\beta}  \right) \quad , \label{eq:2nd-order-tensor}
\end{align}
with:
$$
\begin{array}{rcrl}
  q^{\alpha}           & = &   \lambda & \left( \nabla^{\alpha} T - T U^{\alpha} \partial_{\beta} U^{\beta} \right) \quad , \\
  P^{< \alpha \beta >} & = &   \eta    & \left( \Delta^{\alpha}_{\gamma} \Delta^{\beta}_{\delta} + \Delta^{\alpha}_{\delta} \Delta^{\beta}_{\gamma}  - \frac{2}{3} \Delta^{\alpha \beta} \Delta_{\gamma \delta} \right) \nabla^{\gamma} U^{\delta} \quad , \\
  \varpi               & = & - \mu    & \nabla_{\alpha} U^{\alpha} \quad ; 
\end{array}
$$
$q^{\alpha}$ is the heat flux, $P^{<\alpha \beta>}$ the pressure deviator, $\varpi$ dynamic pressure, $\lambda$ heat conductivity, and $\eta$ and $\mu$ shear and bulk viscosities, respectively. Further we have:
$$
\begin{array}{lcl} 
  \nabla^{\alpha}         & = & \Delta^{\alpha \beta} \partial_{\beta}       \quad ,\\
  \Delta^{\alpha \beta}   & = &  g^{\alpha \beta} - U^{\alpha} U^{\beta}     \quad ,\\
  \Delta^{\alpha}_{\beta} & = & \Delta^{\alpha \gamma} \Delta_{\gamma \beta} \quad .\\
\end{array} 
$$


A kinetic formulation, on the other hand, describes the fluid as a system of interacting particles of 
rest mass $m$; the particle distribution function $f( x^{\alpha}, p^{\beta} )$ 
depends on space-time coordinates $x^{\alpha} = \left( t, \bm{x} \right)$ and momenta 
$p^{\alpha} = \left( p^0, \bm{p} \right) = \left( \sqrt{\bm{p}^2+m^2}, \bm{p} \right)$;
$f(\bm{x}, t, \bm{p}) \diff \bm{x} \diff \bm{p}$ counts the number of particles in the 
corresponding volume element in phase space. 

The system evolves according to the Boltzmann equation, which, 
in the absence of external forces, reads as follows:
\begin{equation}\label{eq:relativistic-boltzmann}
  p^{\alpha} \frac{\partial f}{\partial x^{\alpha}} = \Omega(f) \quad .
\end{equation}
The collision term $\Omega(f)$ is often replaced by simplified models.
For instance, the Anderson-Witting model \cite{anderson-witting-1974a} (a relativistic extension of the well known Bhatnagar-Gross-Krook \cite{bhatnagar-1954} formulation), compatible with the Landau-Lifshitz decomposition, reads
\begin{equation}\label{eq:anderson-witting}
  \Omega = \frac{p^{\mu} U_{\mu}}{\tau} \left( f - f^{eq} \right) \quad .
\end{equation}
The equilibrium distribution $f^{eq}$, following Boltzmann statistics, has been derived many decades ago by J\"uttner \cite{juettner-1911},
\begin{equation}
  f^{eq} \simeq e^{- p^{\mu}U_{\mu}/T} \quad .
\end{equation}
The Anderson-Witting model has just one parameter, the equilibration (proper-)time $\tau$ and
obeys the conservation equations:
\begin{align}
  \partial_{\alpha} N^{\alpha}              &= 0 \quad , \label{eq:cons1}\\
  \partial_{\beta } T^{\alpha \beta}        &= 0 \quad . \label{eq:cons2}
\end{align}

As discussed in previous paragraphs, a predictive bridge between kinetic theory and hydrodynamics must
provide the macroscopic transport coefficients $\lambda, \mu, \eta$, from the mesoscopic ones ($\tau$ in the Anderson-Witting
model). 
Our attempt at contributing further understanding of the issue is based on the following 
analysis; we: i) consider a relativistic flow for which we are able to compute an approximate hydrodynamical 
solution depending on the transport coefficients; ii) study the same flow numerically with a  
lattice Boltzmann kinetic algorithm, obtaining a numerical calibration of the functional
relation between the transport coefficients and $\tau$; iii) obtain clear-cut evidence 
that the CE method successfully matches the numerical results and, iv) double-check our approach 
using the calibrations obtained in ii) for a numerical study of a different relativistic flow, successfully 
comparing with other numerical data obtained by different methods.


We consider Taylor-Green vortices \cite{taylor-green-1937}, a well known example of a 
non-relativistic decaying flow featuring an exact solution of the Navier-Stokes equations, and derive 
an approximate solution in the mildly relativistic regime.
In the non-relativistic case, from the following initial conditions in a 2D periodic domain:
\begin{equation}
  \begin{aligned}\label{eq:tg-classic-initial-conditions}
    u_x(x,y,0) &=& v_0 \cos{\left(x\right)} \sin{\left(y\right)} &, \\
    u_y(x,y,0) &=&-v_0 \cos{\left(y\right)} \sin{\left(x\right)} &, \quad x,y \in [0, 2 \pi]
  \end{aligned}
\end{equation}
the solution is given by
\begin{equation}
  \begin{aligned}\label{eq:tg-classic-solution}
    u_x(x,y,t) &=& v_0 \cos{\left(x\right)} \sin{\left(y\right)} F(t) &, \\
    u_y(x,y,t) &=&-v_0 \cos{\left(y\right)} \sin{\left(x\right)} F(t) &, \quad x,y \in [0, 2 \pi]
  \end{aligned}
\end{equation}
with
\begin{equation}
  F(t) = \exp{\left( -2 \nu ~ t \right)} \quad ,
\end{equation}
where $\nu$ is the kinematic viscosity of the fluid.

In the relativistic case, we need to solve the 
conservation equations (\autoref{eq:cons1}, \autoref{eq:cons2}). 
We consider a system with a constant initial particle density, and assume that density remains 
constant. We will verify later this assumption against our numerical results showing that density fluctuations in time are very small. 
In this case \autoref{eq:cons1} is directly satisfied
and the expression of the second order tensor slightly simplifies, 
since $\nabla_{\alpha} U^{\alpha} = 0$. 
Consequently we drop the term depending on bulk viscosity and rewrite the second order tensor as:
\begin{equation}\label{eq:2nd-order-tensor-simplified}
  T^{\alpha \beta} = - P g^{\alpha \beta} + (\epsilon + P) U^{\alpha} U^{\beta} + P^{< \alpha \beta >}  \quad .
\end{equation} 

We consider the same initial conditions as in \autoref{eq:tg-classic-initial-conditions}, and look for a 
a solution in the form of \autoref{eq:tg-classic-solution}, 
with an appropriate function $F_R(t)$ replacing $F(t)$.
We plug \autoref{eq:tg-classic-solution}
in \autoref{eq:2nd-order-tensor-simplified} and derive bulky analytic expressions for the derivatives of the second order tensor.
A linear expansion of these expressions in terms of $v_0$ yields
a much simpler expression for $\partial_{\beta } T^{\alpha \beta}$, leading to the differential equation
\begin{equation}
  2 \eta F_R(t) + (P + \epsilon) F_R^{'}(t) = 0 \quad .
\end{equation}
Assuming $P + \epsilon$ constant, for a fixed value of $\zeta$, we derive an explicit solution:
\begin{equation}\label{eq:tg-relativistic-approx}
  F_R(t)  = \exp{ \left( - \frac{2 \eta}{P + \epsilon} t  \right) } F_R(0) \quad ,
\end{equation}
depending on just one transport coefficient, the shear viscosity $\eta$.
Observe that while the quantity $P + \epsilon$ exhibits some time variation (as found in the simulations)
due to the evolution of the local temperature, such fluctuations were found to be negligible.

\begin{figure}[htb]
  \centering
  \includegraphics[width=.49\textwidth]{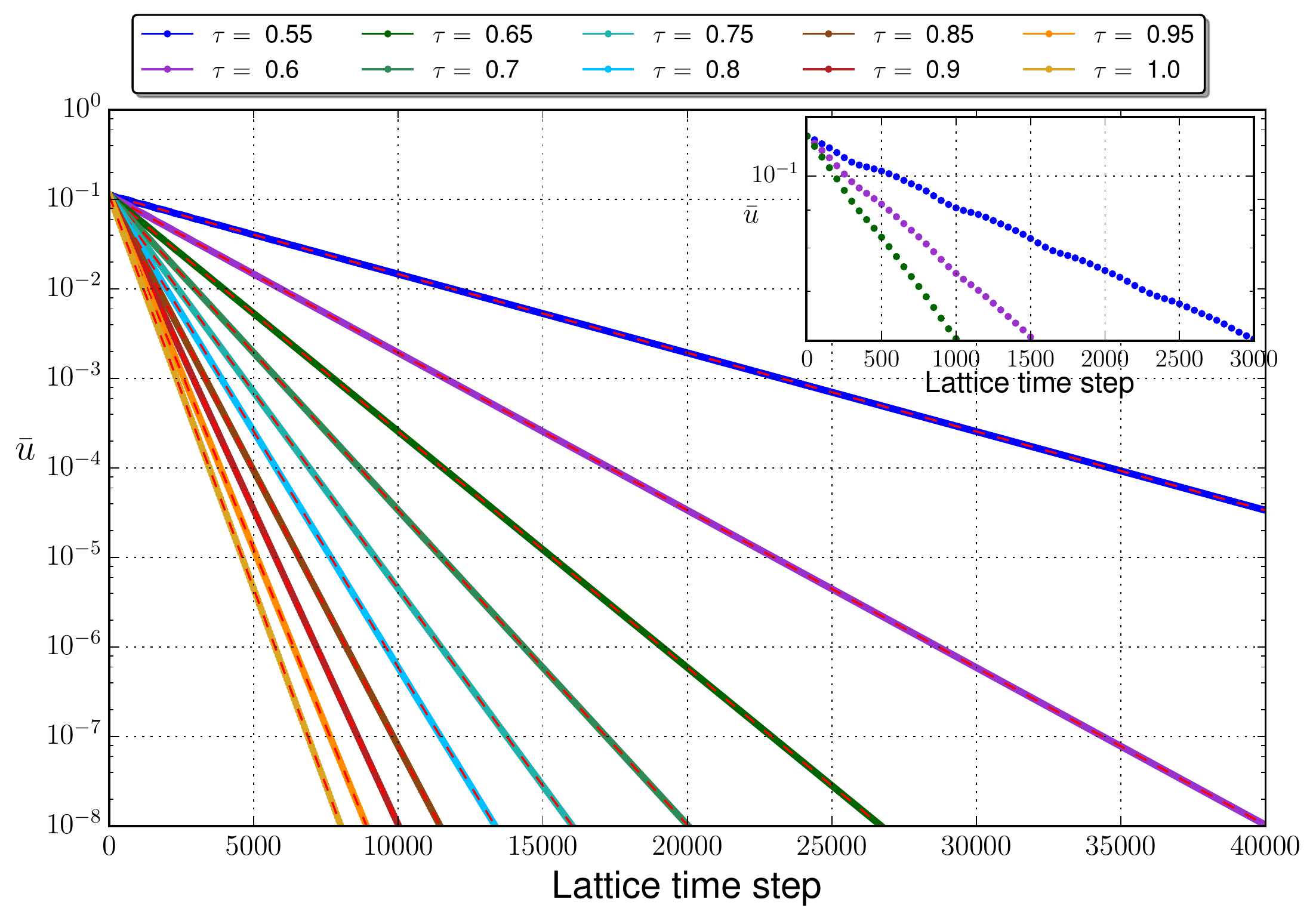}
  \caption{Simulated time evolution of $\bar{u}$ for selected $\tau$ values
  on a $L = 400$ square lattice ($\zeta = 0$, $v_0 = 0.2$, $n_0 = 1$, $T_0 = 1$). Lines are fits to the exponential decay predicted 
  by \autoref{eq:tg-relativistic-approx}. The inset shows non-linear effects in the early phases of the flow.}
\label{fig:m0}
\end{figure}

Next, we compare this analytical solution with data obtained via our LB numerical simulation, aiming 
at linking $\eta$ to the relaxation time $\tau$.
We perform several simulations with 
different values of the initial speed $v_0$ and the mesoscopic parameters, $\tau$ and $\zeta$. 
We consider small (yet, non negligible) values of $u$ and a very broad range of $\zeta$ values, smoothly
bridging between ultra-relativistic to near non-relativistic regimes.
To this end, it is expedient to introduce the observable $\bar{u}$:
\begin{equation}
  \bar{u}^2 = \int \int \left(u_x^2 + u_y^2 \right) \diff x \diff y \quad ,
\end{equation}
defined to be proportional to $F_R(t)$. 
\autoref{fig:m0} gives an example of our numerical results, 
showing the time evolution of $\bar{u}$, clearly exhibiting an exponential decay. 
For each set of mesoscopic values, we perform a linear fit of $\log( \bar{u} )$
extracting a corresponding value for $\eta$ via \autoref{eq:tg-relativistic-approx}.
We next assume a dependence of $\eta$ on the mesoscopic parameters, which, on dimensional 
grounds, reads as 
\begin{equation}\label{eq:rlbm-viscosity}
  \eta = k~f\left( \zeta \right)~ P ~ (\tau - \frac{1}{2})  \quad ,
\end{equation}
with $f(\zeta)$ normalized such that $f(0) = 1$. 
The numerical value of $k$ and the functional form of $f(\zeta)$ contain the physical 
information on the relation between kinetic and hydrodynamics coefficients. 
For instance, CE predicts $k = 4/5$ and an expression for $f(\zeta)$ to which we shall return shortly; for comparison, 
Grad's method predicts $k = 2/3$ and a different functional dependence on $\zeta$.
We are now able to test that \autoref{eq:rlbm-viscosity} holds correctly, checking that all measurements of $\eta(\tau)$ 
at a fixed value of $\zeta$ yield a constant value for $k~f(\zeta)$. 
\begin{table}
\centering 
\begin{tabular}{|c|c|c|c|c|c|c|c|}
\hline 
        & \multicolumn{7}{c|}{$k ~ f(\zeta)$} \\
\hline 
$\tau$  & $\zeta$ = 0 & $\zeta$ = 1.6 & $\zeta$ = 2 & $\zeta$ = 3 & $\zeta$ = 4 & $\zeta$ = 5 & $\zeta$ = 10 \\
\hline
  0.600 & 0.8003 & 0.8319 & 0.8448 & 0.8587 & 0.8892 & 0.8994 & 0.9311  \\
  0.700 & 0.8002 & 0.8318 & 0.8447 & 0.8584 & 0.8888 & 0.8990 & 0.9302  \\
  0.800 & 0.8002 & 0.8318 & 0.8447 & 0.8583 & 0.8887 & 0.8989 & 0.9300  \\
  0.900 & 0.8002 & 0.8318 & 0.8447 & 0.8583 & 0.8887 & 0.8988 & 0.9299  \\
  1.000 & 0.8002 & 0.8317 & 0.8446 & 0.8582 & 0.8887 & 0.8988 & 0.9299  \\
\hline 
\end{tabular}
\caption{Fitted values of $k ~ f(\zeta)$ for selected values of $\tau$ and $\zeta$. Statistical errors are smaller than $1$ in the last displayed digit.}\label{tab:est}
\end{table}
One immediately sees from the second column of \autoref{tab:est} that $k = 4/5$ to very high accuracy, consistently with previous 
results \cite{denicol-2012, tsumura-2012, florkowski-2013, chattopadhyay-2015}. 
More interesting is the assessment of the functional behavior of $f(\zeta)$. 
The CE expansion predicts \cite{cercignani-2002}
\begin{equation}\label{eq:fMT-chapman-enskog}
  f( \zeta ) = \frac{\zeta^3}{12} 
    \left(
    \frac{3}{\zeta^2} \frac{K_3(\zeta)}{K_2(\zeta)} - \frac{1}{\zeta} 
    + \frac{K_1(\zeta)}{K_2(\zeta)} -  \frac{ Ki_1}{K_2(\zeta)}  
    \right) \quad ,
\end{equation}
with $ Ki_1  =  \int_0^{\infty} e^{-\zeta \cosh (t)}/\cosh (t) \textit{dt}$. 

\begin{figure}[htb]
  \centering
  \includegraphics[width=.49\textwidth]{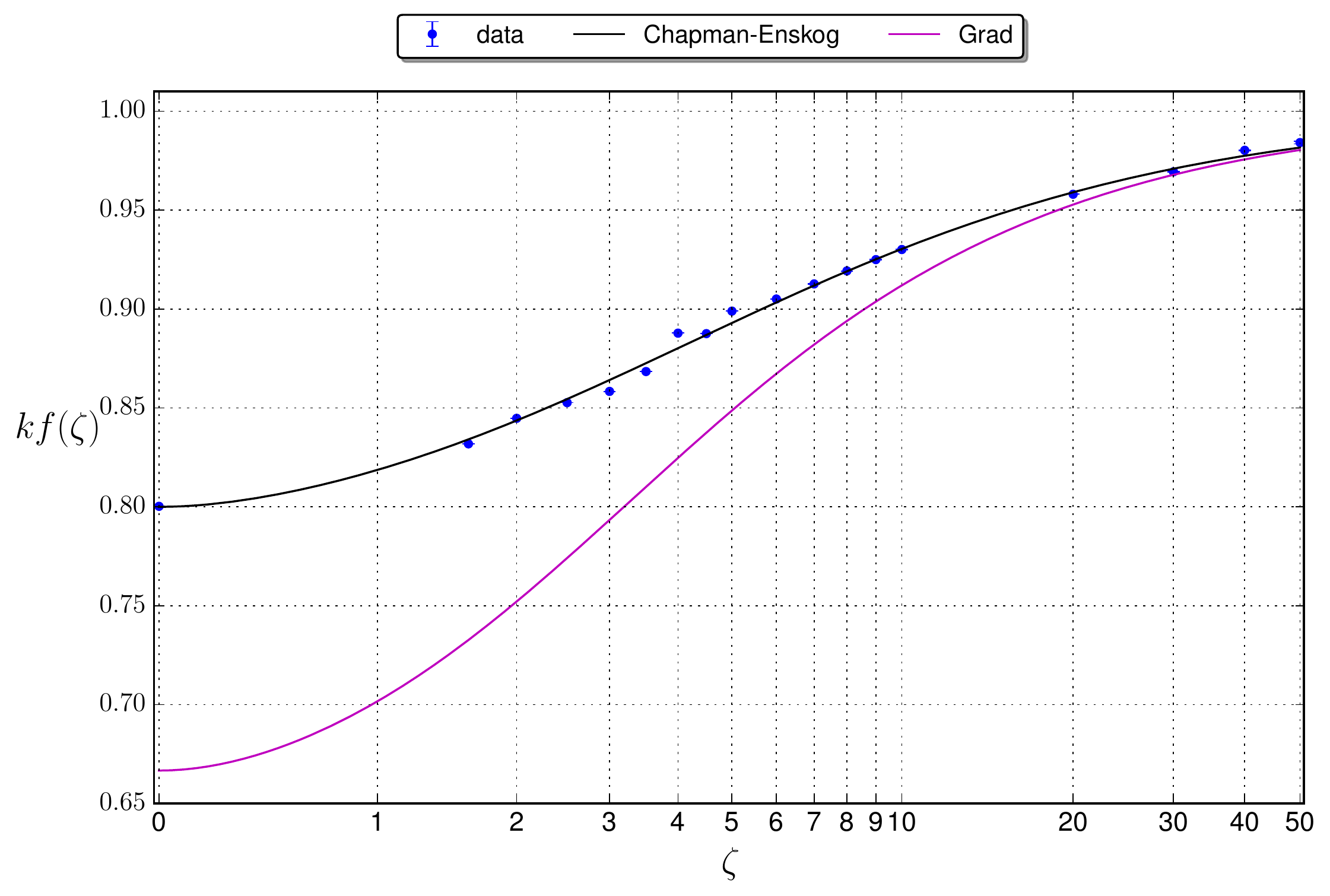}
  \caption{Measured value $k~f( \zeta )$ as a function of $\zeta$. The black (magenta) lines are analytic
           results of the Chapman Enskog (Grad's) methods for the relativistic Boltzmann equation. 
           To improve resolution at small $\zeta$ values,
           we map $\zeta \rightarrow \log{ ( \zeta + \sqrt{1+{\zeta}^2} ) }$ on the x-axis. }
\label{fig:fit-f0-50-o3}
\end{figure}

Our numerical findings for $k~f( \zeta )$ are shown in \autoref{fig:fit-f0-50-o3};
For some $\zeta$ values we have used several different quadratures for our LB
method (see Ref.~\cite{gabbana-2017}), the corresponding results differing from
each other by approximately $1\%$; we consider this an estimate of our systematic
errors. \autoref{fig:fit-f0-50-o3} also shows the CE prediction
(\autoref{eq:fMT-chapman-enskog}) that almost perfectly matches our results (we
remark that {\em no} free parameters are involved in this
comparison) and nicely goes over to the well-known non-relativistic limit for
large values of $\zeta$. For a more quantitative appreciation of
the significance of our result, we also plot the predictions of Grad's method, which
obey the following equation:
\begin{equation}\label{eq:fMT-grad}
  f( \zeta ) = \frac{3}{2} \frac{K_3^2(\zeta)}{K_2(\zeta) ~ K_4(\zeta)} \quad .
\end{equation}
Comparison of the two curves allows to conclude that our level of resolution is adequate to discriminate between the two options.
We performed the same procedure for fully three dimensional simulations, and the corresponding results hold similar degree of accuracy; details will be 
presented in an expanded version of this Letter. 


\begin{figure}[htb]
  \centering
  \includegraphics[width=.49\textwidth]{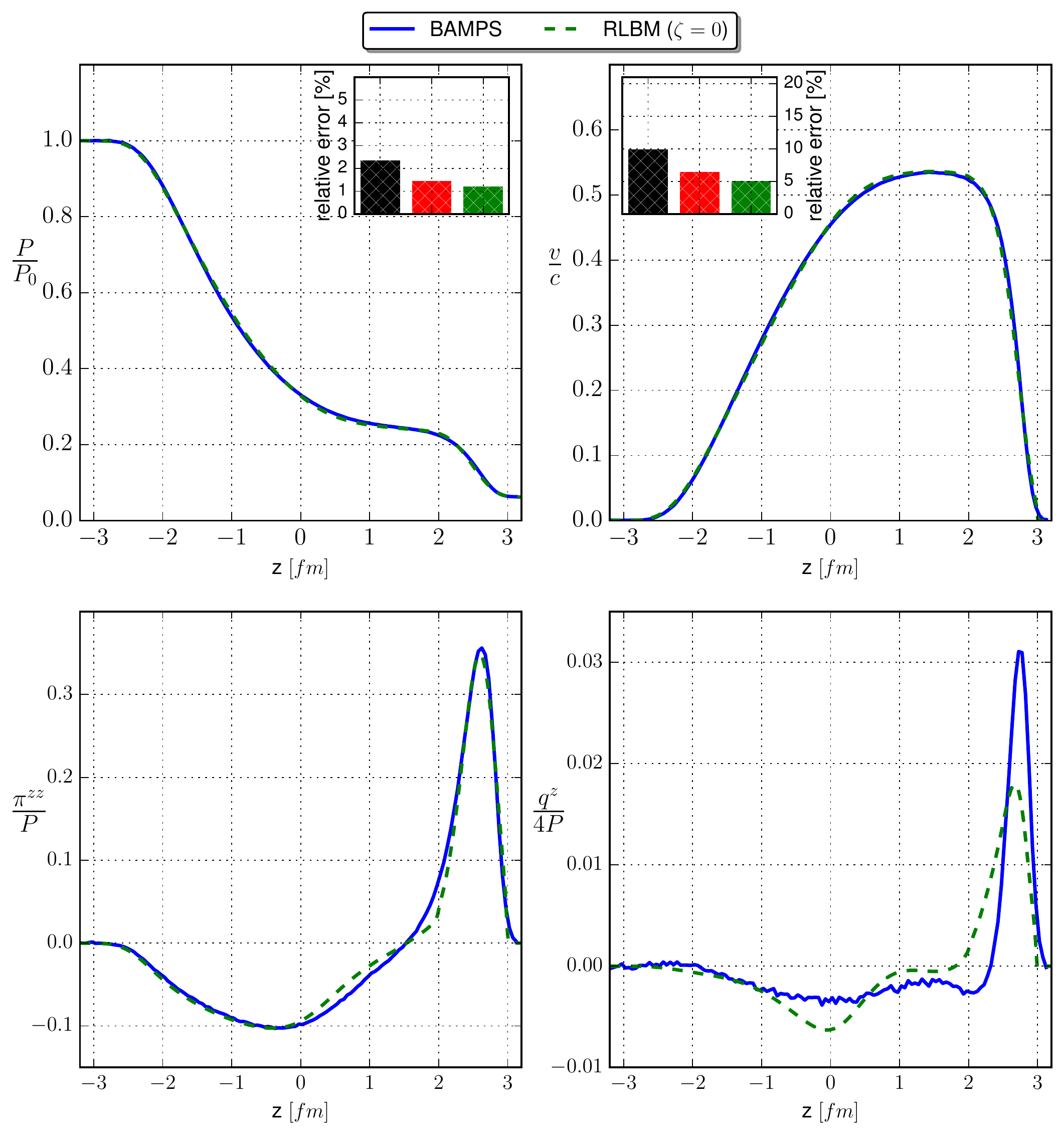}
  \caption{Comparison of BAMPS and our RLBM for the Riemann
           problem at $t = 3.2~fm$ ($\zeta = 0,  \eta/s = 0.1$).
           Top: left) pressure profile; right) velocity profile.
           Error bars are the L2-difference between BAMPS and RLBM using Grad L=1600 (black),
           Chapman Enskog L=1600 (red), Chapman Enskog L=12800 (green).
           Bottom: left) viscous pressure tensor; right) heat flux profile.
           }\label{fig:cmp-bamps-rlbm}
\end{figure} 

Finally, in order to provide a further test of the robustness of our calibration procedure, 
we consider a significantly different problem, 
we simulate a 1D shock tube problem in the ultra-relativistic regime ($\zeta = 0$), 
comparing with BAMPS \cite{xu-2005}, a Monte Carlo numerical solver for the full Boltzmann equation.
This simulation uses a $1 \times 1 \times LZ$ lattice and keeps
the ratio $\eta/s = 0.1$ fixed ($s$ is the entropy density).
The initial conditions for the temperature are $T_A = 400 MeV$ for $ z < 0$
and $T_B = 200 MeV$ for $ z \geq 0$. Initial values for the
pressure step are $P_A = ~5.43 GeV / fm^3$ and $P_B = ~0.339 GeV / fm^3$. 

\autoref{fig:cmp-bamps-rlbm} shows that our results are in excellent agreement with those of BAMPS.
Error bars show the improvement obtained adopting CE for the 
transport coefficients (red bars) over previous results \cite{mendoza-2013} using Grad's method of moments (black bars). 
In \autoref{fig:cmp-bamps-rlbm} we also present the profile of the $\pi^{zz}$ component of
the pressure viscous tensor and of the $q^z$ component of the heat flux, showing good agreement
with results produced by BAMPS for the former quantity, while non-negligible differences arise for the latter.
The reason is that since the Anderson Witting model only provide a free parameter $\tau$, a fine
description of several transport coefficients would require extending it to a multi relaxation time collisional operator.


Summarising, we have investigated the kinetic pathway to dissipative relativistic hydrodynamics by comparing 
lattice kinetic simulations with analytical results based on the Chapman-Enskog  method. 
We find very neat evidence supporting recent theoretical findings in favour of the Chapman-Enskog procedure, which 
we tentatively interpret as the failure of the Grad's method to secure positive-definiteness of the Boltzmann's distribution function.
Since violations of positive-definiteness are most likely to occur in the high-energy tails of the 
distribution, it is natural to speculate that they should be of particular relevance to the 
relativistic hydrodynamic regime, in which tails are significantly more populated than in the non-relativistic case. 
These results are potentially relevant to the study of a wide host of dissipative relativistic hydrodynamic problems, such as 
electron flows in graphene and quark-gluon plasmas \cite{sachdev-2010, mendoza-2011}.
A further intriguing question pertains to the relevance of this analysis to strongly-interacting 
holographic fluids obeying the AdS-CFT bound \cite{maldacena-1999}. 
Indeed, while such fluids are believed to lack a kinetic description altogether, since quasi-particles are 
too short-lived to carry any physical relevance, they are still amenable to 
a {\it lattice} kinetic description, reaching down to values of $\eta/s$ well below the AdS-CFT bound \cite{mendoza-2015, succi-2015}. 
Work to explore the significance of the AdS-CFT bounds in lattice fluids is currently underway.


AG has been supported by the European Union's Horizon 2020 research and
innovation programme under the Marie Sklodowska-Curie grant agreement No. 642069.
MM and SS thank the European Research Council (ERC) Advanced Grant No. 319968-FlowCCS for financial support. 
The numerical work has been performed on the COKA computing cluster at Universit\`a di Ferrara. 


\bibliography{biblio}


\end{document}